\documentclass[twocolumn,showpacs,preprintnumbers,superscriptaddress,amsmath,amssymb]{revtex4}
\usepackage{graphicx}% Include figure files
\usepackage{dcolumn}% Align table columns on decimal point
\usepackage{bm}% bold math

\begin{document}

\preprint{APS/123-QED}

\title{Bulk Screening in Core-Level Photoemission from Mott-Hubbard and Charge-Transfer Systems}

\author{M. Taguchi}

\affiliation{Soft X-ray Spectroscopy Lab, RIKEN/SPring-8, Mikazuki, Sayo, Hyogo 679-5148, Japan}

\author{A. Chainani}

\affiliation{Soft X-ray Spectroscopy Lab, RIKEN/SPring-8, Mikazuki, Sayo, Hyogo 679-5148, Japan}

\author{N. Kamakura}

\affiliation{Soft X-ray Spectroscopy Lab, RIKEN/SPring-8, Mikazuki, Sayo, Hyogo 679-5148, Japan}

\author{K. Horiba}

\affiliation{Soft X-ray Spectroscopy Lab, RIKEN/SPring-8, Mikazuki, Sayo, Hyogo 679-5148, Japan}

\author{Y. Takata}

\affiliation{Soft X-ray Spectroscopy Lab, RIKEN/SPring-8, Mikazuki, Sayo, Hyogo 679-5148, Japan}

\author{M. Yabashi}

\affiliation{Coherent X-ray Optics Lab, RIKEN/SPring-8, Mikazuki, Sayo, Hyogo 679-5148, Japan}
\affiliation{JASRI/SPring-8, Mikazuki, Sayo, Hyogo 679-5148, Japan}

\author{K. Tamasaku}

\affiliation{Coherent X-ray Optics Lab, RIKEN/SPring-8, Mikazuki, Sayo, Hyogo 679-5148, Japan}

\author{Y. Nishino}

\affiliation{Coherent X-ray Optics Lab, RIKEN/SPring-8, Mikazuki, Sayo, Hyogo 679-5148, Japan}

\author{D. Miwa}

\affiliation{Coherent X-ray Optics Lab, RIKEN/SPring-8, Mikazuki, Sayo, Hyogo 679-5148, Japan}

\author{T. Ishikawa}

\affiliation{Coherent X-ray Optics Lab, RIKEN/SPring-8, Mikazuki, Sayo, Hyogo 679-5148, Japan}

\author{S. Shin}

\affiliation{Soft X-ray Spectroscopy Lab, RIKEN/SPring-8, Mikazuki, Sayo, Hyogo 679-5148, Japan}
\affiliation{Institute for Solid State Physics, University of Tokyo, Kashiwa, Chiba 277-8581, Japan}

\author{E. Ikenaga}

\affiliation{JASRI/SPring-8, Mikazuki, Sayo, Hyogo 679-5148, Japan}

\author{T. Yokoya}

\affiliation{JASRI/SPring-8, Mikazuki, Sayo, Hyogo 679-5148, Japan}

\author{K. Kobayashi}

\affiliation{JASRI/SPring-8, Mikazuki, Sayo, Hyogo 679-5148, Japan}

\author{T. Mochiku}

\affiliation{National Institute for Materials Science, Tsukuba, Ibaraki 305-0047, Japan}

\author{K. Hirata}

\affiliation{National Institute for Materials Science, Tsukuba, Ibaraki 305-0047, Japan}

\author{K. Motoya}

\affiliation{Department of Physics, Faculty of Science and Technology, Science University of Tokyo, Noda 278, Japan}

\date{\today} 

\begin{abstract}
We report bulk-sensitive hard X-ray ($h\nu$ = 5.95 keV)
core-level photoemission spectroscopy (PES) of single crystal 
V$_{1.98}$Cr$_{0.02}$O$_{3}$ and the high-$T_c$ cuprate 
Bi$_2$Sr$_{2}$CaCu$_{2}$O$_{8+\delta}$ (Bi2212). V$_{1.98}$Cr$_{0.02}$O$_{3}$
exhibits low binding energy "satellites" to the V $2p$ "main lines" in the metallic phase, which are suppressed in the antiferromagnetic insulator phase. 
In contrast, the Cu $2p$ spectra of Bi2212 do not show temperature dependent features, but a comparison with soft X-ray PES 
indicates a large increase in the $2p^5 3d^9$ "satellites" or $3d^9$ weight in the bulk. Cluster model calculations, including full multiplet structure and a screening channel derived from the coherent band at the Fermi energy, give very satisfactory agreement with experiments.  
\end{abstract}

\pacs{71.30.+h, 74.72.Hs, 78.20.Bh, 79.60.-i}

\maketitle
Core-level photoemission spectroscopy (PES) has played a very important role
in our understanding of the electronic structure
of correlated transition metal (TM) and rare-earth compounds.\cite{kan88} The appearance of strong satellite structure accompanying the main peaks in correlated systems
is well known and systematic variations in the position and intensities of these satellites provide us important clues to their electronic structure.\cite{van81,gun83,fuj84} The inter-atomic 
configuration-interaction approach, using a cluster model or Anderson impurity model , gives a quantitative interpretation for satellite intensities and positions, leading to an accurate description of  
the ground state and excitation spectrum.\cite{van81,gun83,fuj84} In this approach, the physics of TM compounds can be described in terms of a few parameters, namely, 
the $d$-$d$ Coulomb repulsion energy $U$, the charge-transfer energy $\Delta$, the 
ligand $p$-TM $d$ hybridization energy $V$, 
and the core-hole-$d$ electron Coulomb attraction energy $U_{dc}$. 
Zaanen, Sawatzky and Allen\cite{zaa85} proposed a classification scheme for TM compounds
which soon evolved into a paradigm. In this scheme, the band gaps of late TM compounds  
are so-called charge-transfer (CT) type with $U > \Delta$. NiO and CuO are typical
examples of CT insulators while the high-$T_c$ cuprates are
CT insulators driven metallic by doping. In contrast, the 
early TM compounds, with $U < \Delta$ are Mott-Hubbard (MH) systems. 
V$_2$O$_3$, with its alloys, plays the role of a classic MH system displaying
a correlation induced metal-insulator transition.\cite{mcw69,mcw73,cas78} While the old picture
of the MH metal-insulator transition involved a complete collapse or a coalescence
of the lower and upper
MH bands into a single band in the metal phase, 
photoemission studies showed the formation of a well-defined
coherent band at the Fermi level in the presence of remnant MH bands
for a series of correlated oxides\cite{fuj92} and very recently, also for V$_2$O$_3$.\cite{mo03} 
The experimental results are in excellent agreement with calculations using
dynamic mean-field theory (DMFT).\cite{mo03,DMFT} 

In spite of these successes of PES, the surface sensitivity of PES has often led to controversies regarding surface versus bulk electronic structure, and hence, hard X-ray (HX)-PES
is very important and promising.\cite{bra97,woi02}
With the development of high-brilliance  synchrotron
radiation sources, HX-PES with
a resolution of 240 meV at a photon energy of 5.95 keV has recently
become available. The escape depth for Cu and V $2p$ core-level photoelectrons using this photon energy is between $\sim$60-80 \AA,\cite{NIST} significantly higher than that with soft X-ray (SX) photons from a Mg- or Al-K$\alpha$ 
source ($\sim$10 \AA).
Thus, it facilitates a bulk electronic structure investigation of materials.\cite{dal01,kob03,cha04} 

In this work, we study bulk sensitive $2p$ core-level HX-PES ($h\nu = 5.95$ keV) of V$_{1.98}$Cr$_{0.02}$O$_{3}$ and the optimally doped high-$T_c$ cuprate (Bi2212) 
as typical examples of MH and CT systems, respectively. Single crystals of V$_{1.98}$Cr$_{0.02}$O$_{3}$ showed a 
sharp metal-insulator transition at 170 K,\cite{mcw73} while Bi2212 showed 
a superconducting $T_c$ of 90 K.\cite{Li94} HX-PES measurements were performed 
in a vacuum of 1 $\times$ 10$^{-10}$ Torr
at undulator beam line BL29XU, SPring-8\cite{Tamasaku} using a Scienta R4000-10KV electron analyzer. The energy width of incident X-rays was 70 meV, and the total energy resolution, $\Delta$E was set to $\sim$ 0.4 eV. SX-PES (h$\nu$ = 1500 eV) was performed at BL17SU, with  $\Delta$E $\sim$ 0.3 eV.  Sample temperature was controlled
to $\pm 2 $K during measurements. Single crystal V$_{1.98}$Cr$_{0.02}$O$_{3}$ was fractured in-situ at 220K, and measured in a temperature ($T$)
cycle (220 K to 90 K to 220 K) to confirm $T$-dependent changes while Bi2212 was peeled with a scotch tape and measured at room temperature (RT) and 30 K. The Fermi level
(E$_{F}$) of gold was measured to calibrate the energy scale.

\begin{figure}
\includegraphics[scale=.52]{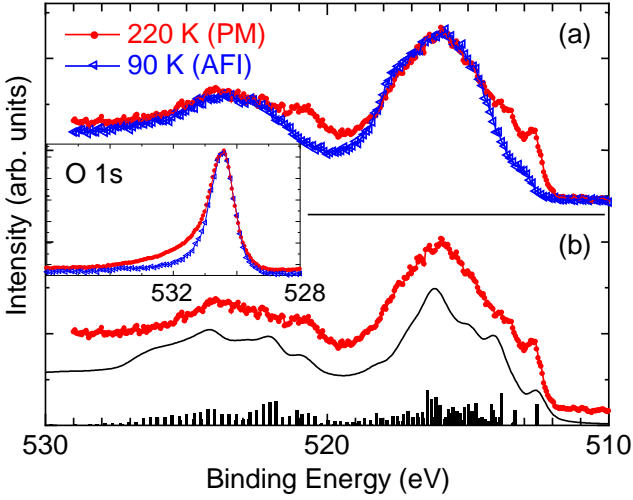}
\caption{\label{fig1}   %FIG.1
(Color online) V $2p$ core-level PES spectra of V$_{1.98}$Cr$_{0.02}$O$_{3}$: (a) Comparison between $T$=220 K (PM phase) and $T$=90 K (AFI phase). (b) Experimental 
V $2p$ PES spectrum 
of V$_{1.98}$Cr$_{0.02}$O$_{3}$ in the PM phase compared with a calculated  spectrum, with an integrated background. Bar diagrams show discrete final states.
Inset shows the 
O $1s$ core-level spectra.}
\vspace{-0.25in}
\end{figure}

The inset to Fig.~1 shows the 
O $1s$ core-level spectra in the paramagnetic metal (PM) and antiferromagnetic insulator (AFI) phases. 
For comparison, the spectrum of AFI phase is shifted by $0.2$ eV to the lower binding energy so as to align it to the PM line. 
The clean single O $1s$ peaks
confirm the high quality of the data. Moreover, in the PM phase, the asymmetry due to electron-hole pair shake-up (the Doniach-\v{S}unji\'c line-shape) is clearly observed while the spectral shape is symmetric in AFI phase. 
In Fig. 1(a) we present the V $2p$ core-level HX-PES spectra measured at 220 K (PM phase)
and 90 K (AFI phase). The spectra consist of the $2p_{3/2}$ and $2p_{1/2}$ 
spin-orbit split features.  A clear change in the  PM phase as compared to the AFI phase,  with a sharp additional feature at 512.5 eV, and structures around 514 eV and
521 eV binding energy are observed. These features are low binding energy
"satellites" to the "main peaks" of  the $2p_{3/2}$ and $2p_{1/2}$ 
spin-orbit split features. The observations of $T$-dependent changes in
O $1s$ and V $2p$ core-levels confirm the metal-insulator transition  in 
V$_{1.98}$Cr$_{0.02}$O$_{3}$.\cite{smith94} The sharp peak 
at 512.5 eV and the  low binding energy satellites we observe here, 
appeared as weak shoulders to the main peak in the earlier study of
 V$_{2-x}$Cr$_{x}$O$_{3}$
using SX-PES, possibly due to the lower resolution and/or the higher surface sensitivity.
Its
origin was tentatively attributed to a difference in core-hole screening between the
metallic and insulating states.\cite{smith94} These $T$-dependent "well-screened"
features cannot be interpreted in 
the usual cluster model or Anderson impurity model applied to
TM compounds since the calculations
do not include a temperature dependent modification of the
$d$-derived states. Further,  
since recent valence band PES of V$_{2}$O$_{3}$ 
shows a prominent coherent peak at the E$_{F}$\cite{mo03} 
which gets gapped in the AFI,\cite{smith94,shin} and the low binding energy "satellites" in V $2p$ core-levels are also observed only in the PM phase but suppressed in the AFI phase, we felt it important to check the possibility 
of screening by states at the E$_{F}$.\cite{Cox}

\begin{figure}
\includegraphics[scale=.45]{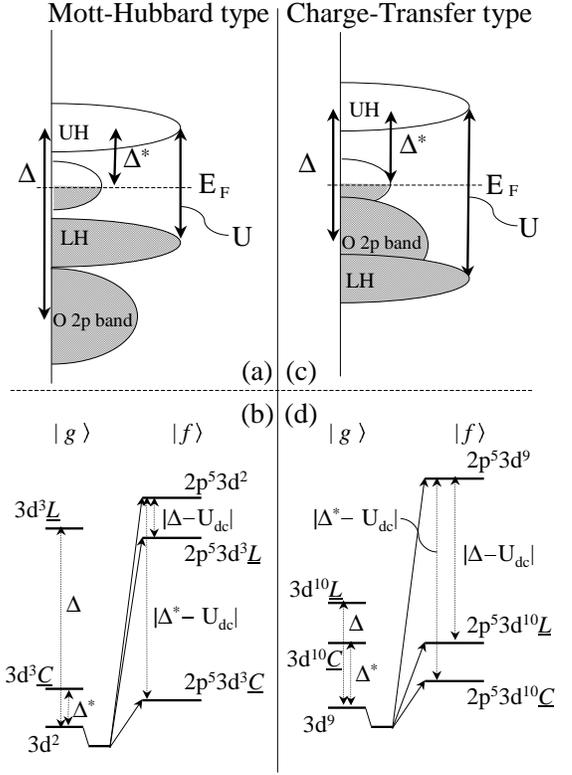}
\caption{\label{fig2}     %FIG.2
Schematic illustration of energy levels and total energy level diagram of $2p$ core-level PES for MH and CT systems in metallic phase. }
\vspace{-0.25in}
\end{figure}

To do so, one may introduce charge transfer from a coherent band at E$_{F}$ within the framework of a cluster model. Here we retain only a single V ion (VO$_{6}$) 
cluster and allow charge transfer between the V site and the ligand sites as well as
the V site and the coherent band, approximated as a level.\cite{imer87,boc95} 
The charge transfer from coherent band can be directly related to the metallic screening in core-level PES, originally proposed by Kotani and Toyozawa\cite{kot74}. 
 As shown 
schematically in Fig.~2(a), the charge transfer energy from coherent band to upper Hubbard (UH) band is $\Delta^*$ whereas the usual charge transfer energy $\Delta$ (from O $2p$ ligand band to UH band),
is defined as the energy difference of the configuration-averaged energies $E(3d^3\underline{L})-E(3d^2)$. The inclusion of states where electrons have been transferred to the V site from the coherent band is expected to describe the low binding energy satellites in the metallic phase. 

Numerical calculations were carried out based on the configuration-interaction cluster model with intra-atomic full multiplets in C$_{3v}$ local symmetry and including the screening channel 
for charge transfer from the coherent band. 
The ground state is described by a linear combination of following configurations: $3d^2$, $3d^3\underline{L}$, $3d^4 \underline{L}^2$, $3d^1 C$, $3d^3 \underline{C}$, $3d^4 \underline{L}\underline{C}$, and $3d^4 \underline{C}^2$,  where $C$ represents the electron in the coherent band just above E$_{F}$, $\underline{C}$ is the hole state in the coherent band just below E$_{F}$ and $\underline{L}$ is the hole state in O $2p$ ligand band. 
 The final states are thus described by a linear combination of $2p^53d^2$, $2p^53d^3\underline{L}$, $2p^53d^4 \underline{L}^2$, $2p^53d^1 C$, $2p^53d^3 \underline{C}$, $2p^53d^4 \underline{L}\underline{C}$, and $2p^53d^4 \underline{C}^2$.
The energy differences of each configuration in the initial and final states are listed in Table~I. 
The Hamiltonian is given by 

\begin{eqnarray}
H &=&  H_{I}+H_{II}, \\
   \nonumber \\ 
H_{I} &=&  \sum_{\Gamma,\sigma} \varepsilon_{3d}(\Gamma)d^{\dagger}_{\Gamma\sigma}d_{\Gamma\sigma}
 \hspace{-1pt}+ \hspace{-1pt}\sum_{m,\sigma}\varepsilon_{2p}p^{\dagger}_{m \sigma}p_{m \sigma} \nonumber \\
 &+& \sum_{\Gamma,\sigma}\varepsilon_{p}(\Gamma)a^{\dagger}_{\Gamma\sigma}a_{\Gamma\sigma}
 + \sum_{\Gamma,\sigma}V(\Gamma)(d^{\dagger}_{\Gamma\sigma}a_{\Gamma\sigma}
 + a^{\dagger}_{\Gamma\sigma}d_{\Gamma\sigma}) \nonumber \\
 &+& U_{dd}\sum_{(\Gamma,\sigma)\neq(\Gamma',\sigma')}d^{\dagger}_{\Gamma\sigma}
d_{\Gamma\sigma}d^{\dagger}_{\Gamma'\sigma'}d_{\Gamma'\sigma'}\hspace{2.0cm} \nonumber \\
   &-& U_{dc}(2p)\sum_{\Gamma,m,\sigma,\sigma'}d^{\dagger}_{\Gamma\sigma}d_{\Gamma\sigma}
(1 - p^{\dagger}_{m\sigma'}p_{m\sigma'})
\nonumber \\
   &+& H_{\rm{multiplet}}, \\
   \nonumber \\ 
H_{II} &=&  \sum_{\Gamma,\sigma} \varepsilon_{c}(\Gamma)c^{\dagger}_{\Gamma\sigma}c_{\Gamma\sigma} \nonumber \\
 &+& \sum_{\Gamma,\sigma}V^*(\Gamma)(d^{\dagger}_{\Gamma\sigma}c_{\Gamma\sigma} + c^{\dagger}_{\Gamma\sigma}d_{\Gamma\sigma}).  \ \
\end{eqnarray}

\noindent
The first term $H_I$ of the total Hamiltonian $H$ represents the standard cluster model.\cite{tag97} In addition to the usual cluster model ($H_I$ term), we have introduced states labeled '$C$' responsible for a new screening effect described by $H_{II}$ term in Eq. (1). These new states represent the doping-induced states which develop into a metallic band at E$_F$.  $\varepsilon_{3d}(\Gamma)$, $\varepsilon_{2p}$, $\varepsilon_{p}(\Gamma)$ and $\varepsilon_{c}(\Gamma)$ represent the energies of V $3d$, V $2p$, O $2p$ ligand states and doping-induced states at E$_{F}$, respectively, with the irreducible representation (= $a_1$, $e_g^{\sigma}$, and $e_g^{\pi}$) of the $C_{3v}$ symmetry. The indices m and $\sigma$ are the orbital and spin states. $V(\Gamma)$, $U_{dd}$, and $-U_{dc}(2p)$ are the hybridization between V $3d$ and O $2p$ ligand states, the on-site repulsive Coulomb interaction between V $3d$ states and the attractive $2p$ core-hole potential, respectively. The Hamiltonian $H_{\rm{multiplet}}$ describes the intra-atomic multiplet coupling originating from the multipole components of the Coulomb interaction between V $3d$ states and that between V $3d$ and $2p$ states. The spin-orbit interactions for V $2p$ and $3d$ states are also included in $H_{\rm{multiplet}}$. 
Note that our Hamiltonian is essentially the same as that of Bocquet $et$ $al.$\cite{boc95}, but with an additional intra-atomic multiplet interaction term.
More significantly, there is an important difference in the basis set used by Bocquet $et$ $al.$ and the present work. In the present work, we use additional new basis states of the type $3d^{n+m} \underline{C}^m$ to account for screening from doping-induced states, while Bocquet $et$ $al.$ have used the basis set to consist of $3d^{n+m}\underline{L}^m$ and $3d^{n-m}C^m$ states to describe the ground state of Ni compounds.

  An effective coupling parameter for describing the interaction strength between the central V $3d$ orbitals and the coherent band, $V^* (\Gamma)$, is introduced analogous to the hybridization $V(\Gamma)$. We allowed the $3d$-band hybridization to be reduced by a factor $R_c$ $(=0.8)$ in the presence of core-hole and enhanced by a factor $1/R_v$ $(=1/0.9)$ in the presence of an extra $3d$ electron.\cite{gun88} Following the recent analysis of linear dichroism, we also include a small negative trigonal crystal field $D_{trg}$.\cite{par00,tan02} 

\begin{table}
\begin{tabular}{ccccccccc}
\multicolumn{4}{c}{}\\
\hline
$\text{Configuration}$ & $\text{initial state}$ & $\text{final state}$  & \\
\hline
 $E(3d^3\underline{L})-E(3d^2)$ & $\Delta$ & $\Delta - U_{dc}$  & \\
 $E(3d^4\underline{L}^2)-E(3d^2)$ & $2\Delta+U_{dd}$ & $2\Delta +U_{dd} - 2U_{dc}$  & \\
 $E(3d^1C)-E(3d^2)$ & $U_{dd}-\Delta^*$ & $U_{dd}-\Delta^* + U_{dc}$  & \\
 $E(3d^3\underline{C})-E(3d^2)$ & $\Delta^*$ & $\Delta^*-U_{dc}$  & \\
 $E(3d^4\underline{C}^2)-E(3d^2)$ & $2\Delta^*+U_{dd}$ & $2\Delta^*+U_{dd}-2U_{dc}$  & \\
 $E(3d^4\underline{L}\underline{C})-E(3d^2)$ & $\Delta+\Delta^*+U_{dd}$, & $\Delta+\Delta^*+U_{dd}-2U_{dc}$  & \\
\hline

\end{tabular}
\caption{Energy differences for each configurations in both initial and final stats in $2p$ core-level PES.}
%\vspace{-0.25in}
\end{table}

Figure 1(b) shows our theoretical spectrum for PM phase, 
compared to the experimental spectrum at 220 K. 
We used the following parameters for the C$_{3v}$ cluster : $U_{dd}=4.5$, $\Delta=6.0$, $U_{dc}=6.5$, $10Dq=1.2$, $D_{trg}=-0.05$, $V(e_g^{\sigma})=2.9$, $\Delta^*=0.9$, $V^*(e_g^{\sigma})=0.75$, in units of eV. For checking the validity of the estimated parameter sets, we have also calculated the linear dichroism for the same parameter sets and obtain a good agreement with previous results.\cite{par00,tan02} 
Thus, theory and experiment show very satisfactory agreement for the
complete multiplet structure and the low binding energy satellites.  
Note that, in the limit of $V^*(\Gamma) \to 0$, our cluster model reduces to the conventional single cluster model and calculated spectrum is identical to the previous theory for SX-PES\cite{zim98}.

To clarify the peak assignment, the total energy level diagram is shown in Fig.~2(b) for V$_{2}$O$_{3}$. 
The ionic configurations are used in the absence of hybridization and multiplet terms. The $3d^4 \underline{L}^2$, $3d^1C$, $3d^4 \underline{C}^2$ and $3d^4 \underline{C} \underline{L}$ configurations are not depicted for simplicity.  Since $\Delta^*$ is smaller than $\Delta$, the $3d^3\underline{C}$ state lies just above $3d^2$ ones in the initial state. 
As for final states, $2p^53d^3\underline{L}$ have energies around the $2p^53d^2$, 
whereas the $2p^53d^3\underline{C}$ state lies clearly below them. As a consequence, the main lines are due to a mixture of $2p^53d^3\underline{L}$ and $2p^53d^2$, whereas the low binding energy satellites are mainly due to the coherently screened $2p^53d^3\underline{C}$ final states .

Since the screening from states at E$_F$ imply long range or non-local screening, we felt it important to make a comparison with the high-$T_c$ cuprate (Bi2212) as a CT system. 
The Cu $2p$ spectra of the cuprates is complex and extensive 
work has shown the role of non-local and local screening in explaining the data obtained using SX-PES.\cite{vee93,koi02}
 Figure~3(a) shows the Cu $2p_{3/2}$ HX-PES spectra of Bi2212 (filled circles) compared with 
SX-PES data (open triangles) obtained for the same sample at RT. 
The 30 K spectrum of HX-PES (open circles) is also shown in Fig.~3(a). 
The HX-PES spectra do not show significant $T$-dependent changes.
But the SX : HX comparison clearly shows that the spectral weight in the high binding energy 
satellite increases significantly in the bulk when we normalize at $933$ eV binding energy.  It is noted that the escape depth of  
$\sim$10 $\text{\AA}$  in SX-PES\cite{NIST} probes only top two Cu-O layers (c-parameter $\approx $  30 \AA)
while the present HX-PES probes at least 2-3 unit cells. 
The large increase in intensity ($\sim$ 50\%) of the total spectral weight in HX-PES compared to SX-PES
would naively suggest increase of $3d^9$ weight in the bulk 
since the satellite is generally assigned to the $2p^53d^9$ state.

\begin{figure}
\includegraphics[scale=.46]{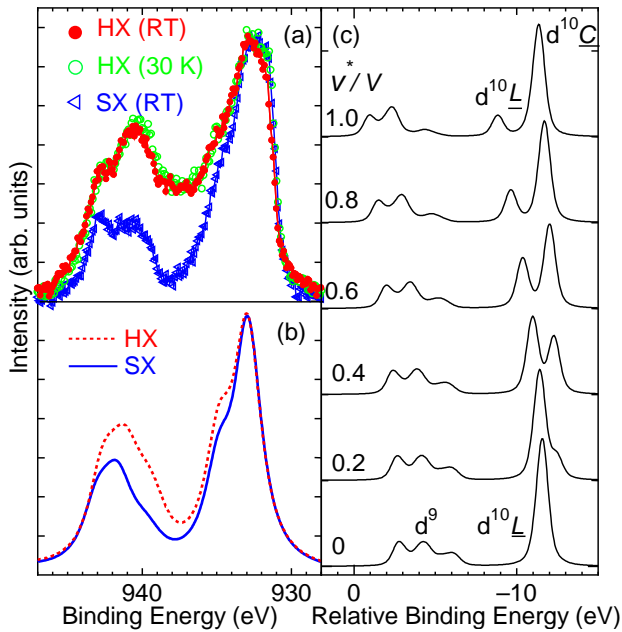}
\caption{\label{fig3}   %FIG.3
(Color online) Cu $2p_{3/2}$ of Bi2212. (a) Comparison between HX (5.95 keV) (filled circles) and SX (1.5 keV) (open triangles) at RT and with $T=30$ K (open circles) HX-PES after subtracting an integral background. (b) D$_{4h}$ cluster calculations for the HX-PES (dashed line) and SX-PES (solid line) spectra.  (c) $V^*$ dependence of Cu $2p_{3/2}$ PES. The other parameter values are fixed to the values for SX-PES stated in the text.}
\vspace{-0.25in}
\end{figure}

Calculated results of Cu $2p_{3/2}$ PES are also shown in Fig.~3(b). We use the same model as  in the case of V$_{1.98}$Cr$_{0.02}$O$_{3}$ but change the local symmetry to D$_{4h}$. 
The parameter values used are as follows: (i) for HX-PES $U_{dd}=7.2$, $\Delta=2.75$, $U_{dc}=9.0$, $T_{pp}=1.0$ (the hybridization between nearest neighbor O $2p$ orbitals), $V(e_g)=2.0$, $V^*(e_g)=1.62$, and $\Delta^*=1.75$; (ii) for SX-PES $U_{dd}=7.2$, $\Delta=2.75$, $U_{dc}=9.0$, $T_{pp}=1.0$, $V(e_g)=2.5$, $V^*(e_g)=1.87$, and $\Delta^*=1.75$ in the unit of eV. 
The calculated results are in good qualitatively agreement with experiment, except for the width 
or structure in the main peak at $932$ eV. The structure in the main peak originates in the valence band structure and/or Zhang-Rice singlet formation, as is known from Anderson impurity model or multi-site calculations,\cite{koi02} beyond the present model. 
The fitted parameter values indicate that hybridization $V$ and $V^*$ are reduced in the bulk HX-PES compared to SX-PES. 
This is somewhat surprising as it implies a decrease of the hybridization strength in the bulk. In general, the different atomic environment and
reduced co-ordination, often conspire to reduce hybridization and screening at the surfaces. Further experimental and theoretical studies are necessary to clarify this issue. 

A schematic energy diagram for Bi2212 is shown in Fig.~2(d). Similar to the case of V$_2$O$_3$, the $3d^9$ state gives the biggest contribution to the ground state. Since this is a CT type system (i.e. $U > \Delta$), 
the core-hole potential pulls down both the $2p^53d^{10}\underline{C}$ and $2p^53d^{10}\underline{L}$ states which lie below the $2p^53d^9$ state. The $2p^53d^{10}\underline{C}$ is the lowest energy state but its energy is very close to $2p^53d^{10}\underline{L}$ state. Therefore the lowest binding energy 
peak at 933 eV in the calculation is due to $2p^53d^{10}\underline{C}$ while the broad feature at 935 eV is due to the locally screened peak denoted by "$2p^53d^{10}\underline{L}$". The 933 eV feature can be identified with non-local screening 
effect.\cite{vee93,koi02} 
To confirm the $3d^{10}\underline{C}$ state as the non-local screening peak, we calculate the $V^*$ dependence of Cu $2p_{3/2}$ PES spectra as shown in Fig.~3(c). The calculated spectra without $3d^{10}\underline{C}$ has a single $3d^{10}\underline{L}$ peak and is identical to the results 
of single ion cluster model calculation. When the hybridization $V^*$ is switched on, the 
$3d^{10}\underline{C}$ state appears and grows in intensity for increasing $V^*$. 
This behavior is identical to the non-local screening effect.\cite{vee93,koi02} A recent study reported use of DMFT to calculate core-level spectra, 
but in the absence of ligand states and multiplet structure\cite{kim04}. 
They have concluded that the low binding energy satellites observed
 in a series of Ru-oxides, which display metal-insulator transition, can also
be consistently explained in terms of a coherent screening channel at E$_{F}$.

Since we use HX-PES with a photon energy of $\sim$ 6 keV, it is also important to discuss the possibility of (i) multi-pole effect ($i.e.$ break down of the dipole approximation),\cite{tse78} (ii) double photo-excitation effect\cite{arm85} as an origin for the spectral changes observed by us. Since both (i) and (ii) are known to be atomic in origin, they are expected not to exhibit temperature or doping dependence as observed for V$_{2-x}$Cr$_{x}$O$_{3}$, as well as for another system La$_{1-x}$Sr$_{x}$MnO$_{3}$ (LSMO) studied recently by HX-PES\cite{hor04}. In both these systems, coherent screening accounts very well for the well-screened low binding energy feature. Furthermore, we have also checked the probing depth dependence by changing the photon energy and emission angle for LSMO\cite{hor04}. The results indicate that the surface effect component is minimized for the largest probing depth geometry as used in the present HX-PES study.

Finally, from a general viewpoint, it should be emphasized that the metallic screening mechanism discussed here can be related to Kotani and Toyozawa model as was originally applied to elemental metals.\cite{kot74} In TM compounds, the original metallic screening has been ignored in core-level SX-PES because the spectra never showed a metallic screening feature even in the metal phase. The ligand screening was found to be enough to explain the spectra. The reason why conventional core-level PES showed no big difference between metal and insulating phase remains to be answered and the present study using hard X-ray provides an answer to this long standing issue using the probing depth variation with photon energy. The present model thus shows the importance of metallic screening effects in addition to ligand screening effects.

In summary, core-level HX-PES was used to investigate V$_{1.98}$Cr$_{0.02}$O$_{3}$
and optimally doped Bi2212 as examples of MH and CT systems. 
V$_{1.98}$Cr$_{0.02}$O$_{3}$ displays clear changes in the 
O $1s$ and V $2p$ spectral shapes across the metal-insulator transition. From a configuration-interaction cluster model analysis, the low binding energy satellite is assigned to bulk screening. 
In contrast to V$_{1.98}$Cr$_{0.02}$O$_{3}$, the Cu $2p$ core-level of Bi$_2$Sr$_{2}$CaCu$_{2}$O$_{8+\delta}$ 
shows significant increase in the $2p^5 3d^9$ "satellite" intensity in HX-PES compared to
SX-PES, suggesting an increase of the $3d^9$ weight in the bulk. 
The lowest binding energy features in MH and CT type correlated metals exhibit bulk screening from the coherent band. The model is also shown to reproduce the non-local screening peak of 
multi-site or Anderson impurity model calculations, making the model suitable for wide applications.

We gratefully acknowledge valuable discussions with Prof. Akio Kotani and Prof. Kozo Okada.

\end{document}